# Modeling Urban Growth and Form with Spatial Entropy

Yanguang Chen

(Department of Geography, College of Urban and Environmental Sciences, Peking University, Beijing 100871, P.R. China. E-mail: chenyg@pku.edu.cn)

**Abstract**: Entropy is one of physical bases for fractal dimension definition, and the generalized fractal dimension was defined by Renyi entropy. Using fractal dimension, we can describe urban growth and form and characterize spatial complexity. A number of fractal models and measurements have been proposed for urban studies. However, the precondition for fractal dimension application is to find scaling relations in cities. In absence of scaling property, we can make use of entropy function and measurements. This paper is devoted to researching how to describe urban growth by using spatial entropy. By analogy with fractal dimension growth models of cities, a pair of entropy increase models can be derived and a set of entropy-based measurements can be constructed to describe urban growing process and patterns. First, logistic function and Boltzmann equation are utilized to model the entropy increase curves of urban growth. Second, a series of indexes based on spatial entropy are used to characterize urban form. Further, multifractal dimension spectrums are generalized to spatial entropy spectrums. Conclusions are drawn as follows. Entropy and fractal dimension have both intersection and different spheres of application to urban research. Thus, for a given spatial measurement scale, fractal dimension can often be replaced by spatial entropy for simplicity. The models and measurements presented in this work are significant for integrating entropy and fractal dimension into the same framework of urban spatial analysis and understanding spatial complexity of cities.

# 1. Introduction

In recent years, urban research is gradually evolving into a new science, that is, urban science based on ideas from complexity theory. It is now known that cities and regions are complex spatial systems, which cannot be modeled and analyzed by using conventional mathematical methods (Albeverio *et al*, 2008; Allen, 1997; Batty, 2005; Portugali, 2011; Wilson, 2000). Spatial complexity can be described with both fractal dimension and entropy (Bar-Yam, 2004a; Bar-Yam, 2004b; Batty, 2005; Chen, 2008; Wilson, 2000; White and Engelen, 1994). If an urban phenomenon bears characteristic scales, we can characterize it by means of entropy, and if the urban phenomenon has no characteristic scale, we have to depict it by using fractal dimension (Batty and Longley, 1994; Chen and Huang, 2018; Shen, 2002; Sun and Southworth, 2013; Tannier *et al*, 2011; Thomas *et al*, 2010). However, in practice, things are usually more complicated and more difficult for spatial systems such as urban morphology and urban traffic networks. It is easy to describe a simple system or the simple aspect of a complex system. To describe a complex system, we must reveal the scaling relationships behind observed data. Sometimes, the scaling property is covered. In this case, we have to find alternative ways for scientific description. An effective approach to making spatial analysis for cities is to make use of zonal systems (Batty, 1976; Batty and Longley, 1994; Feng, 2002; He *et al*, 2019; Wang and Zhou, 1999; Yang *et al*, 2019). There are three types of zonal systems. The first is natural systems of zones, the second is the artificially defined systems of zones, and the third is the results of recursive subdivision of space. Natural zonal systems come from physical evolution of geographical landscape, artificial zonal system originates from administrative or management partition (Batty, 1976; He *et al*, 2019; Kaye, 1989; Thomas *et al*, 2007), and the recursive subdivision of space is one of means of spatial disaggregation (Batty and Longley, 1994; Goodchild and Mark, 1987). Districts and counties of cities represents artificially defined systems of zones, and box-counting method represents a way of spatial recursive subdivision. Based on the zonal systems of districts and counties, spatial entropy can be calculated for urban phenomena (Batty, 1974; Batty, 1976), and using box-counting method, we can compute both entropy and fractal dimension of urban land use (Chen *et al*, 2017; Chen and Feng, 2017; Feng and Chen, 2010).

Both fractal dimension and entropy can be employed to explore spatial complexity of cities. Fractal geometry has been adopted to research cities, many models are built to reflect urban evolution, and lots of fractal measurements are presented to describe urban properties (Batty and Longley, 1994; Benguigui *et al*, 2000; Chen, 2020; Frankhauser, 1994; Frankhauser, 1998). Among various fractal methods, the most powerful one may be multifractal scaling, which provides effective characterization for spatial complexity of cities (Appleby, 1996; Ariza-Villaverde *et al*, 2013; Chen, 2008; Chen and Wang, 2013; Murcio *et al*, 2015). One of theoretical bases of multifractal theory is Renyi entropy. Based on multifractals and entropy concept, a number of fractal parameters and indexes have been put forward to characterize urban form (Chen, 2020). Based on fractal dimension sets, a set of models are employed to depict urban growth (Chen, 2012; Chen, 2018). In theory, the spatial entropy increase curves of urban evolution should be modeled by sigmoid functions by analogy with fractal dimension growth models. If so, we can describe urban growth by spatial entropy in absence of fractal parameters. However, research report on spatial entropy increase modeling of cities has not yet been found in literature. This paper is devoted to deriving models and measurements based on entropy function by analogy with fractal models of urban growth. The principal goal of this study is to provide a set of mathematical tools for describing urban growth and thus understanding spatial complexity of cities. In Section 2, a set of mathematical models of urban growth are derived from urban fractal dimension increase models, a series of urban indexes are constructed using entropy values. In Section 3, an empirical studies on the national capital city of China are made to testify the models and measurements. In Section 4, several related questions are discussed, and finally, the discussion are concluded by summarizing the main points of this study. The models and measurements developed in this paper are useful for integrating spatial entropy and fractal dimension into the same logic framework of urban spatial analysis.

## 2. Models and measures

### 2.1 Relation between entropy and fractal dimension

The starting point of this study is general entropy, based on which the generalized fractal dimension has been defined for multifractal theory. Multifractal geometry is one of important tools for exploring spatial complexity of cities. The generalized correlation dimension can also be regarded as generalized information dimension, which is on the basis of Renyi's entropy (Renyi, 1961). Based on the box-counting method, the Renyi entropy can be expressed as

$$M_q(\varepsilon) = -\frac{1}{q-1}\ln\sum_{i=1}^{N}P_i(\varepsilon)^q = \ln(\sum_{i=1}^{N}P_i^q)^{1-q} = \ln C_q(\varepsilon), \quad (1)$$

where $q$ represents the order of moment ($q=\ldots,-2,-1, 0, 1, 2,\ldots$), $\varepsilon$ denotes the linear size of boxes for spatial cover and measurement, $N$ is the number of nonempty boxes ($i=1, 2,\ldots, N$), $P_i$ is probability ($\sum P_i=1$), $M_q$ refers to the $q$th order Renyi entropy, $C_q$ is the corresponding generalized correlation function, and the symbol "ln" denotes natural logarithm function. If $q=0$, we will have

$$M_0(\varepsilon) = \ln N, \quad (2)$$

which denotes Hartley macro state entropy; If $q=1$, according to the L'Hospital rule, equation (1) change to

$$M_1(\varepsilon) = -\sum_{i=1}^{N}P_i(\varepsilon)\ln P_i(\varepsilon), \quad (3)$$

which refers to Shannon information entropy; If $q=2$, we will have

$$M_2(\varepsilon) = -\ln\sum_{i=1}^{N}P_i(\varepsilon)^2 = \ln C_2(\varepsilon), \quad (4)$$

which denotes the second order Renyi entropy. Equation (4) displays the relationships between entropy and correlation function. In fact, correlation analysis is a useful approach to estimating fractal dimension (Batty and Longley, 1994; Chen, 2008; Thomas *et al*, 2007; Thomas *et al*, 2008). Correlation functions proved to be significant mathematical methods for characterize urban form (Makse *et al*, 1995; Makse *et al*, 1998).

The entropy function is usually employed to describe the development state and property of the systems with characteristic lengths. For a simple system or the simple aspect of a complex system, the entropy value can be uniquely determined. On the contrary, for the complex systems such as fractal cities, which bear no characteristic length in many aspects, the Renyi entropy is not definite quantity. An effective solution to this problem is to transform the generalized entropy into generalized fractal dimension. Based on Renyi's entropy, the generalized correlation dimension is always defined as follows (Feder, 1988; Grassberger 1985; Mandelbrot, 1999)

$$D_q = -\frac{M_q(\varepsilon)}{\ln\varepsilon} = -\lim_{\varepsilon\to 0}\frac{1}{q-1}\frac{\ln\sum_{i}^{N(\varepsilon)}P_i(\varepsilon)^q}{\ln(1/\varepsilon)} = \lim_{\varepsilon\to 0}\frac{\ln(\sum_{i}^{N(\varepsilon)}P_i(\varepsilon)^q)^{1-q}}{\ln(1/\varepsilon)} = \frac{\ln C_q(\varepsilon)}{\ln(1/\varepsilon)}, \quad (5)$$

where "lim" denotes the limit function in mathematical analysis. Three special fractal dimension parameters were derived from the general definition of fractal dimension (Grassberger, 1983). If the moment order $q=0$, we have capacity dimension, $D_0$, which is given by

$$D_0 = -\frac{M_0(\varepsilon)}{\ln \varepsilon} = -\frac{\ln N}{\ln \varepsilon}, \qquad (6)$$

The capacity dimension corresponds to Hartley entropy; If $q=1$, we obtain information dimension, $D_1$, which is defined by

$$D_1 = -\frac{M_1(\varepsilon)}{\ln \varepsilon} = \frac{1}{\ln \varepsilon} \sum_{i=1}^{N} P_i(\varepsilon) \ln P_i(\varepsilon), \qquad (7)$$

The information dimension corresponds to Shannon information entropy; If $q=2$ as given, then we get correlation dimension, $D_2$, which is expressed as

$$D_2 = -\frac{M_2(\varepsilon)}{\ln \varepsilon} = -\frac{\ln C_2(\varepsilon)}{\ln \varepsilon}, \qquad (8)$$

The correlation dimension corresponds to the second order Renyi entropy. The relationships between Renyi entropy, generalized correlation function, general dimension, and spatial size of boxes are as below:

$$M_q(\varepsilon) = \ln C_q(\varepsilon) = D_q \ln \varepsilon, \qquad (9)$$

$$C_q(\varepsilon) = \exp(M_q(\varepsilon)) = (\sum_{i}^{N(\varepsilon)} P_i(\varepsilon)^q)^{1-q} = \varepsilon^{-D_q}. \qquad (10)$$

This suggests that spatial correlation bears no characteristic scale, but spatial entropy has characteristic scale. According to equation (9), fractal dimension is the characteristic values of spatial entropy. According to equation (10), the fractal dimension is the scaling exponent of spatial correlation function. For a given measurement size $\varepsilon$, spatial entropy is linearly proportional to fractal dimension. Therefore, under certain conditions, the models of fractal dimension increase of cities can be extended to spatial entropy increase of complex spatial systems.

## 2.2 Simple model: logistic function

First of all, the logistic model of spatial entropy increase of urban form can be derived from the fractal dimension growth model of urban form verified by observation evidence. This model is easy to understand. If the minimum dimension $D_{min}=0$, fractal dimension curve of urban growth can be modeled by the logistic function (Chen, 2012; Chen, 2018)

$$D_q(t) = \frac{D_{max}}{1+(D_{max}/D_q(0)-1)e^{-rt}}, \tag{11}$$

where $D_q(t)$ denotes the generalized fractal dimension of moment order $q$ at time $t$, $D_{max}$ is the capacity parameter indicative of the maximum fractal dimension in the future, $D_q(0)$ is the initial value of the fractal dimension at time $t=0$ in the past, $r$ refers to the initial growth rate of fractal dimension. In the case of $D_{min}=0$, the ratio of fractal dimension $D_q(t)$ to the maximum dimension $D_{max}$ proved to equal the ratio of generalized entropy $M_q(t)$ to the maximum entropy $M_{max}$ (Chen, 2020), that is,

$$\frac{D_q(t)}{D_{max}} = \frac{M_q(t)}{M_{max}}. \tag{12}$$

Equation (12) can be converted into the following form

$$D_q(t) = \frac{D_{max} M_q(t)}{M_{max}}. \tag{13}$$

If $t=0$ as given, then equation (12) changes to

$$\frac{D_{max}}{D_q(0)} = \frac{M_{max}}{M_q(0)}, \tag{14}$$

in which $M_q(0)$ is the initial value of the generalized entropy at time $t=0$. Substituting equations (13) and (14) into equation (11) yields

$$D_{max}\frac{M_q(t)}{M_{max}} = \frac{D_{max}}{1+(M_{max}/M_q(0)-1)e^{-rt}}, \tag{15}$$

By eliminating $D_{max}$ in equation (15), we get the logistic model of spatial entropy increase of cities:

$$M_q(t) = \frac{M_{max}}{1+(M_{max}/M_q(0)-1)e^{-rt}}, \tag{16}$$

where $r$ represents the initial growth rate of generalized entropy of urban form.

Based on normalized variable of spatial entropy, the three-parameter logistic model can be simplified as a two-parameter logistic model. If $M_{min}=0$, then equation (16) can also be "normalized" and re-expressed as

$$M_q^*(t) = \frac{M_q(t)}{M_{max}} = \frac{1}{1+(1/M_q^*(0)-1)e^{-rt}}, \tag{17}$$

where $M_q^*(0)=M_q(0)/M_{max}$ is the normalized result of $M_q(0)$ when $M_{min}=0$. The normalized spatial entropy can be used to define the degree of spatial equilibrium. It proved to equal normalized fractal dimension, which is also treated as the index of space filling extent. That is, we have

$$J_q(t) = M_q^*(t) = D_q^*(t), \quad (18)$$

where $J_q(t)$ denotes the degree of spatial equilibrium. A spatial entropy odds can be defined as

$$O_q(t) = \frac{M_q(t)}{I_q(t)} = \frac{M_q(t)}{M_{max} - M_q(t)}. \quad (19)$$

in which $O_q(t)$ refers to the spatial entropy odds, $I_q(t)=M_{max}-M_q(t)$ denotes the information gain (Batty, 1974; Batty, 1976). The maximum entropy $M_{max}$ divided by information gain $I_q(t)$ yields redundancy $Z_q(t)=1- M_q(t)/M_{max}$ (Batty, 1974). Substituting equation (16) into equation (19) yields

$$O_q(t) = \frac{M_q(t)}{I_q(t)} = \frac{\frac{1}{1+(1/M_q^*(0)-1)e^{-rt}}}{1-\frac{1}{1+(1/M_q^*(0)-1)e^{-rt}}} = (1/M_q^*(0)-1)e^{rt} = O_q(0)e^{rt}, \quad (20)$$

where $O_q(0)=M_q(0)/(M_{max}-1)$. This suggests that spatial entropy odds of urban form increases exponentially. Using equation (20), we can define a logit transform of spatial entropy.

## 2.3 Exquisite model: Boltzmann equation

Further, we can derive Boltzmann model of spatial entropy increase of urban form from the corresponding fractal dimension model. In fact, logistic model can be regarded as the special case of Boltzmann equation, which has been used to describe urban growth (Benguigui *et al*, 2001). If the minimum dimension $D_{min}>0$, fractal dimension increase curve of urban form can be modeled by Boltzmann's equation (Chen, 2012)

$$D_q(t) = D_{min} + \frac{D_{max} - D_{min}}{1+(\frac{D_{max} - D_q(0)}{D_q(0) - D_{min}})e^{-rt}} = D_{min} + \frac{D_{max} - D_{min}}{1+\exp(-\frac{t-\tau}{p})}, \quad (21)$$

where $D_{min}$ denotes the lower limit of fractal dimension, $D_{max}$ is the upper limit of fractal dimension, $p$ is a scaling parameter associated with the initial growth rate $r$, and $\tau$, a temporal translational parameter indicative of a critical time, when the growth rate of fractal dimension reaches its peak (Chen, 2018). The scale translation and scaling parameters can be defined respectively by $\tau=\ln((D_{max}-D_q(0))/(D_q(0)-D_{min}))^p$ and $p=1/r$. For $D_{min}>0$, the normalized fractal dimension proved to equal the normalized entropy (Chen, 2020), that is

$$\frac{D_q(t)-D_{\min}}{D_{\max}-D_{\min}}=\frac{M_q(t)-M_{\min}}{M_{\max}-M_{\min}}, \tag{22}$$

which can be re-expressed as

$$D_q(t)-D_{\min}=(D_{\max}-D_{\min})\frac{M_q(t)-M_{\min}}{M_{\max}-M_{\min}}. \tag{23}$$

If $t=0$, then equation (22) can be changed to

$$\frac{D_{\max}-D_{\min}}{D_q(0)-D_{\min}}=\frac{M_{\max}-M_{\min}}{M_q(0)-M_{\min}}. \tag{24}$$

Inserting equations (23) and (24) into equation (21) yields

$$(D_{\max}-D_{\min})\frac{M_q(t)-M_{\min}}{M_{\max}-M_{\min}}=\frac{D_{\max}-D_{\min}}{1+(\frac{M_{\max}-M_{\min}}{M_q(0)-M_{\min}}-1)e^{-rt}}. \tag{25}$$

By eliminating the fractal dimension range, $D_{\max} - D_{\min}$, in equation (25), we have

$$M_q(t)=M_{\min}+\frac{M_{\max}-M_{\min}}{1+(\frac{M_{\max}-M_q(0)}{M_q(0)-M_{\min}})e^{-rt}}=M_{\min}+\frac{M_{\max}-M_{\min}}{1+\exp(-\frac{t-\tau}{p})}, \tag{26}$$

which represent the Boltzmann equation of spatial entropy increase of urban growth. For the normalized variable, equation (26) can be re-written as

$$M_q^*(t)=\frac{M_q(t)-M_{\min}}{M_{\max}-M_{\min}}=\frac{1}{1+(1/M_q^*(0)-1)e^{-rt}}, \tag{27}$$

in which $M_q^*(0)=(M_q(0)-M_{\min})/(M_{\max}-M_{\min})$ denotes the normalized result of $M_q(0)$, the initial value of spatial entropy. Obviously, equation (27) is equivalent in form to equation (17). Due to equation (9), $M_{\max}=D_{\max}\ln\varepsilon$, $M_{\min}=D_{\min}\ln\varepsilon$. From equation (22) it follows

$$D_q(t)=\frac{D_{\max}-D_{\min}}{M_{\max}-M_{\min}}M_q(t), \tag{28}$$

which can be used to generalize multifractal spectrums to Renyi entropy spectrums.

## 2.4 New spatial measurements based on fractal dimension and entropy

Fractal dimension can be treated as a basic measure of urban growth, and this measure is used to replace urban area. Due to scale-dependence of urban spatial measurements, urban area cannot be objectively determined, while fractal dimension is a scale-free parameter, which can be employed to substitute urban area to reflect space filling and land use extent. Based on fractal dimension of urban form, a set of urban measurements or indexes can be defined to describe city development. The measurements based on fractal dimension are tabulated as follows (Table 1). (1) *Fractal dimension range*, the difference between the upper limit and lower limit of fractal dimension values, $D_{max}$-$D_{min}$. (2) *Space-filling degree*, the difference between the fractal dimension value at time *t* and the lower limit of fractal dimension value, $D(t)$-$D_{min}$. (3) *Space-saving degree*, or *space-remaining degree*, the difference between the upper limit of fractal dimension value and the fractal dimension value at time *t*, $D_{max}$-$D(t)$. (4) *Space-filling ratio*, the ratio of space-filling degree to fractal dimension range, $(D(t)$-$D_{min})/(D_{max}$-$D_{min})$. (5) *Space-saving ratio*, or *space-remaining ratio*, the ratio of space-saving degree to fractal dimension range, $(D_{max}$-$D(t))/(D_{max}$-$D_{min})$. (6) *Fractal dimension odds*, the ratio of the fractal dimension value at time *t* to space-saving degree, $D(t)/(D_{max}$-$D(t))$. The basic relationships between these indexes are as below: (a) The space-filling degree plus space-saving degree equals fractal dimension range; (b) The space-filling ratio plus space-saving ratio equals 1; (c) If $D_{min}$=0, then the space-filling degree divided by space-saving degree equals fractal dimension odds.

**Table 1 A set of spatial measurements based on fractal dimension of urban growth and form**

| Measurement (fractal index) | Definition ($D_{min}>0$, $D_{max}<d_E$) | Special case 1 ($D_{min}=0$) | Special case 2 ($D_{min}=0$, $D_{max}=d_E$) | Meaning |
|---|---|---|---|---|
| Fractal dimension range | $D_{max} - D_{min}$ | $D_{max}$ | $d_E$ | Available space |
| Space-filling degree | $D(t) - D_{min}$ | $D(t)$ | | Used space |
| Space-saving degree | $D_{max} - D(t)$ | | $d_E - D(t)$ | Remained space |
| Space-filling ratio | $\dfrac{D(t) - D_{min}}{D_{max} - D_{min}}$ | $\dfrac{D(t)}{D_{max}}$ | $\dfrac{D(t)}{d_E}$ | Proportion of used space |
| Space-saving ratio | $\dfrac{D_{max} - D(t)}{D_{max} - D_{min}}$ | $\dfrac{D_{max} - D(t)}{D_{max}}$ | $\dfrac{d_E - D(t)}{d_E}$ | Proportion of remained space |
| Fractal dimension odd | $\dfrac{D(t)}{D_{max} - D(t)}$ | | $\dfrac{D(t)}{d_E - D(t)}$ | Ratio of used space to remained space |

All these new measurements based on fractal dimension of cities can be generalized to spatial entropy of complex spatial systems. The space-filling ratio proved to be a normalized fractal dimension, and the normalized fractal dimension proved to equal the normalized spatial entropy of urban form (Chen, 2012; Chen, 2020). The spatial entropy reflects the land-use extent of an urban region, namely, the degrees of space-filling and spatial uniformity. The measurements based on spatial entropy are tabulated as below (Table 2). (1) *Spatial entropy range*, the difference between the upper limit and lower limit of spatial entropy values, $M_{max}$-$M_{min}$. (2) *Space-filling index*, the difference between the spatial entropy value at time $t$ and the lower limit of spatial entropy value, $M(t)$-$M_{min}$. (3) *Space-saving index*, or *spatial information gain*, the difference between the upper limit of spatial entropy value and the spatial entropy value at time $t$, $I(t)=M_{max}$-$M(t)$. (4) *Space-filling ratio*, the ratio of space-filling index to spatial entropy range, $(M(t)$-$M_{min})/(M_{max}$-$M_{min})$. If $M_{min}=0$ as given, then we have *spatial entropy ratio* $M(t)/M_{max}$. (5) *Space-saving ratio*, or *space-remaining ratio*, the ratio of space-saving index or spatial information gain to spatial entropy range, $(M_{max}$-$M(t))/(M_{max}$-$M_{min})$. If $M_{min}=0$ as given, then we have *spatial redundancy* $Z(t)=1$-$M(t)/M_{max}$. (6) *Spatial entropy odds*, the ratio of the spatial entropy value at time $t$ to the space-saving index or spatial information gain, $M(t)/(M_{max}$-$M(t))$. The basic relationships between these indexes are as follows: (a) The space-filling index plus space-saving index equals spatial entropy range; (b) The space-filling ratio plus space-saving ratio equals 1; (c) If $D_{min}=0$, then the space-saving index divided by the space-filling index equals spatial entropy odds. (d) 1 minus spatial entropy ratio equals *spatial redundancy*.

**Table 2 A set of spatial measurements based on entropy of urban growth and form**

| Measurement (entropy index) | Definition | Special case 1 ($M_{min}=0$) | Special case 2 ($M_{min}=0$, $M_{max}=\ln N$) | Meaning |
|---|---|---|---|---|
| Spatial entropy range | $M_{max} - M_{min}$ | $M_{max}$ | $\ln N$ | Available space |
| Space-filling index | $M(t) - M_{min}$ | $M(t)$ | | Used space |
| Space-saving index; Spatial information gain | $M_{max} - M(t)$ | | $\ln N - M(t)$ | Remained space |
| Space-filling ratio; Spatial entropy ratio | $\dfrac{M(t) - M_{min}}{M_{max} - M_{min}}$ | $\dfrac{M(t)}{M_{max}}$ | $\dfrac{M(t)}{\ln N}$ | Proportion of used space, degree of spatial equilibrium |

| Space-saving ratio; Spatial Redundancy | $\dfrac{M_{max} - M(t)}{M_{max} - M_{min}}$ | $\dfrac{M_{max} - M(t)}{M_{max}}$ | $\dfrac{\ln N - M(t)}{\ln N}$ | Proportion of remained space |
| --- | --- | --- | --- | --- |
| Spatial entropy odds | $\dfrac{M(t)}{M_{max} - M(t)}$ | | $\dfrac{M(t)}{\ln N - M(t)}$ | Ratio of remained space to used space |

In fact, two of the above entropy indexes are widely used in various scientific fields. One is information gain, and the other is redundancy (Batty, 1974; Batty, 1976; Batty, 2010). Prigogine once proposed an *H*-quantity based on Ehrenfest urn and Shannon entropy (Prigogine and Stengers, 1984). The *H*-quantity can be employed to describe spatial difference of urban form. Suppose an urban area is divided into *N* parts, and each part represent a zone. The formula of *H*-quantity is as below:

$$H_t = \sum_{k=1}^{N} P(k,t) \ln \frac{P(k,t)}{P_{eqm}(k)}, \qquad (29)$$

where $H_t$ is the *H*-quantity at time *t*, $P(k, t)$ denotes the probability of the *k*th zone at time *t*, and $P_{eqm}(k)$ represents the probability of absolute uniform distribution (the number *k*=1, 2, 3, …, *N*). For the *N* zones, we have $P_{eqm}(k)=1/N$. Thus, the *H*-quantity can be re-expressed as

$$\begin{aligned} H_t &= \sum_{k=1}^{N} P(k,t) \ln \frac{P(k,t)}{P_{eqm}(k)} = \sum_{k=1}^{N} P(k,t)[\ln P(k,t) - \ln P_{eqm}(k)] \\ &= -M(t) - \sum_{k=1}^{N} P(k,t) \ln P_{eqm}(k) = \sum_{k=1}^{N} P(k,t) \ln N - S(t) \\ &= \log N - M(t) = M_{max} - M(t) \end{aligned} \qquad (30)$$

where $M(t)$ refers to Shannon information entropy. This suggests that the *H*-quantity is in fact information gain (Batty, 1974; Batty, 2010). That is, $H_t = I(t)$. The *H*-quantity divided by the maximum entropy yields spatial redundancy (Batty, 1974)

$$Z(t) = \frac{H_t}{M_{max}} = 1 - \frac{M(t)}{M_{max}}. \qquad (31)$$

The entropy of time *t* divided by the *H*-quantity yields spatial entropy odds

$$O = \frac{M(t)}{H_t} = \frac{M(t)}{M_{max} - M(t)}. \qquad (32)$$

This suggests that the *H*-quantity can act as a basic measurement of urban growth and form. Please note that the measurements such as *H*-quantity, information gain and redundancy in literature are based on Shannon entropy (Batty, 1974; Batty, 1976; Prigogine and Stengers, 1984). In this work, they are extended to general expressions based on Renyi entropy.

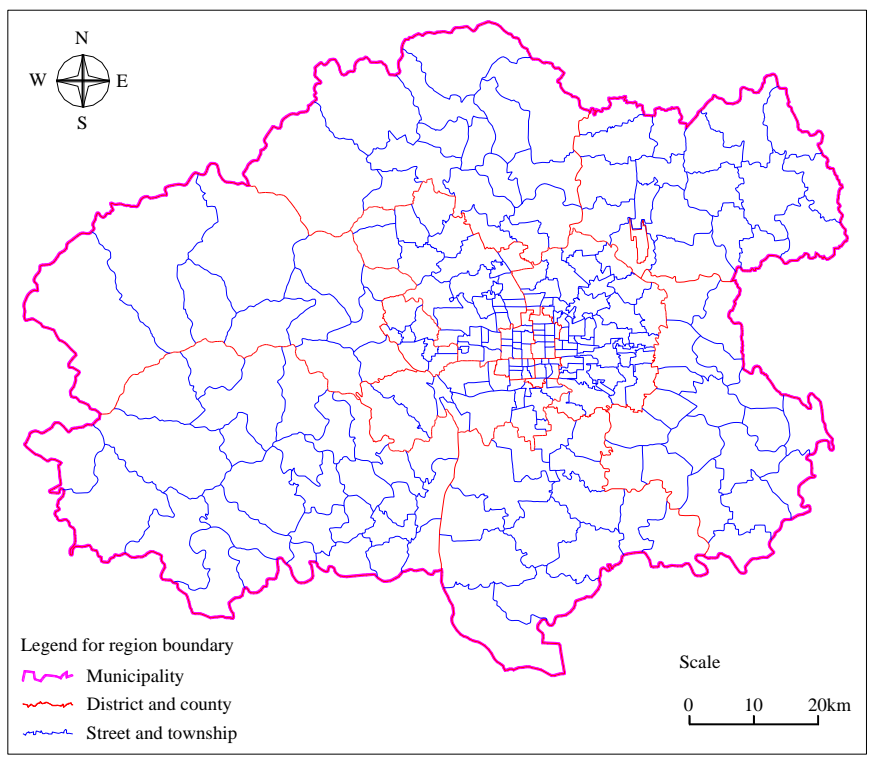

**Figure 1 A zonal system of Beijing city with two levels: districts and residential area (2010)**
**Note**: Beijing Municipality is divided into 18 districts and counties (thick red lines indicate boundaries). A district contains several streets, and a county contains several townships and towns (thin blue lines represent boundaries). There are 331 streets, towns, and townships in Beijing region. Thus, the zonal system of districts and counties include 18 units, and the zonal system of residential areas include 331 units.

## 3. Empirical analysis

### 3.1 Modeling urban growth of Beijing city

A simple example is presented here to show how to utilize the models and entropy-based indexes developed in this paper to characterize urban growth and form. The national capital city of China, Beijing, can be employed to make empirical analyses (Figure 1). The original data proceeded from remote sensing images of Beijing city from 1984 to 2015. Huang (2019) extracted these data and computed the fractal dimension of Beijing's urban form. The properties of the methods lie in three aspects. First, the spatial sphere of study area is fixed. Second, the functional box-counting method is adopted to extracted data. Third, the regression analysis based the least squares methods is used to evaluate parameters (Huang and Chen, 2018). Using the final datasets, I can calculate the spatial entropy based on different measurement scales. The spatial entropy based on the minimum scale which reaches the resolution limit is the desirable (Table 3). With these spatial entropy values, we can make two studies: one is verify the logistic model of spatial entropy increase, and the other, is to characterize the spatial complexity of Beijing city using the entropy values. What is more, we can draw a comparison between the spatial measurements based on entropy and those based on fractal dimension.

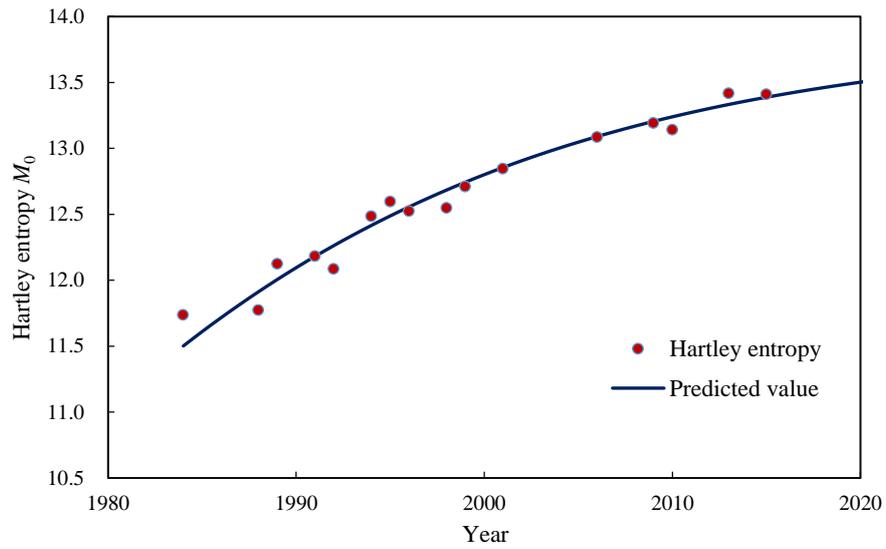

a. Hartley entropy

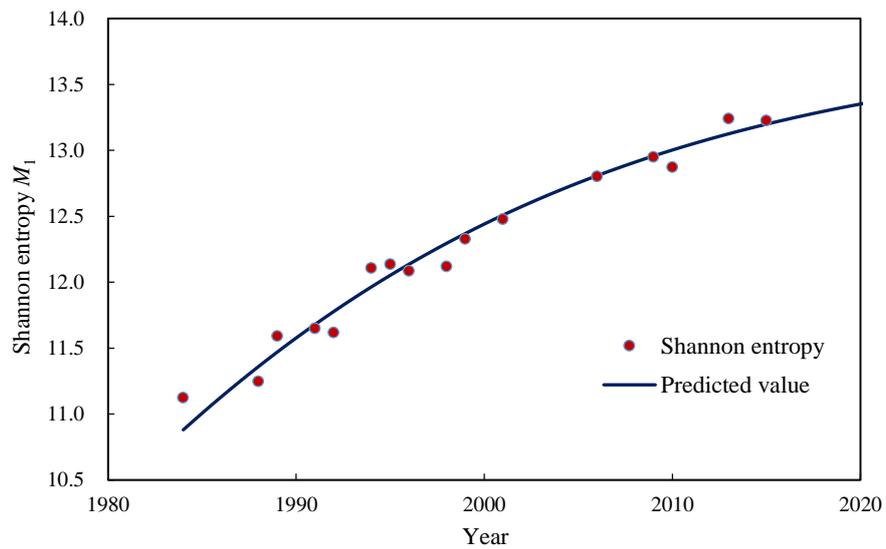

b. Shannon entropy

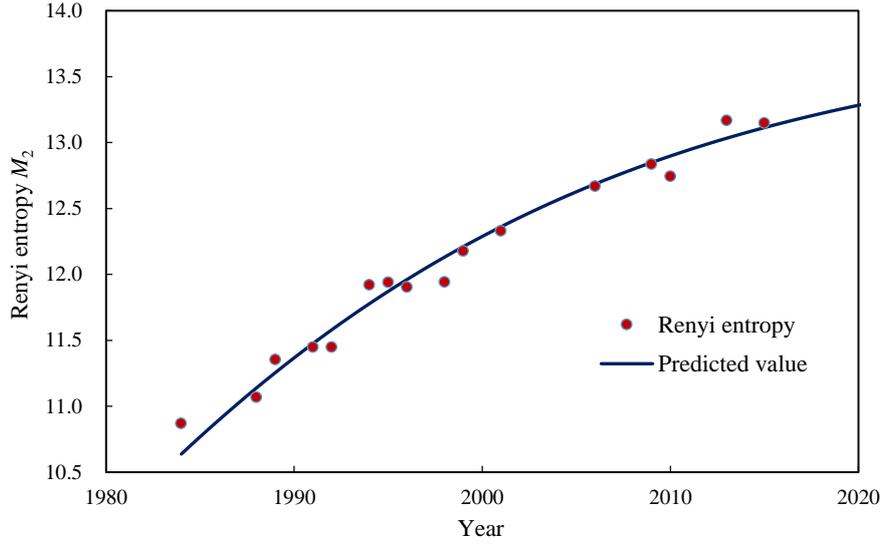

c. The 2nd order Renyi entropy

**Figure 2 The time series of spatial entropy of Beijing's urban land use patterns (1984-2015)**
**Note**: The observed values come between 1984 and 2015. The plots showed the predicted values before 2020.

The typical entropy measures can be utilized to make positive analysis of logistic models of spatial entropy increase. Among Renyi entropy spectrum, there are three basic measures, that is, Hartley entropy $M_0$, Shannon entropy $M_1$, and the second order Renyi entropy $M_2$. They correspond to capacity dimension $D_0$, information dimension $D_1$, and correlation dimension $D_2$, in multifractal dimension spectrum. For $q=0$, we have Hartley entropy. By using the least square calculation based on the observed data, we can build a model as follows

$$\hat{M}_0(t) = \frac{13.8629}{1 + 0.2053 e^{-0.0566t}}, \qquad (33)$$

where $t=n-1984$, and $n$ represent year. The goodness of fit is $R^2=0.9709$, and the estimated initial value of Hartley entropy is $M_0(0)=11.5013$. For $q=1$, we have Shannon information entropy. Using the least square method, we make a model as below

$$\hat{M}_1(t) = \frac{13.8629}{1 + 0.2741 e^{-0.0547t}}. \qquad (34)$$

The coefficient of determination is $R^2=0.9728$, the estimated initial value of Shannon entropy $M_1(0)$ =10.8805. For $q=2$, we have the second order Renyi entropy. Using the least squares calculation, we construct a model as follows

$$\hat{M}_2(t) = \frac{13.8629}{1 + 0.3032 e^{-0.0538t}}. \qquad (35)$$

The multiple correlation coefficient squared is $R^2=0.9733$, the estimated initial value of the 2nd order Renyi entropy is $M_2(0) = 10.6373$. By means of these models, we can predicted the spatial entropy increase of Beijing's city evolution in future (Figure 2).

**Table 3 Spatial entropy values and corresponding fractal dimension of urban form of Beijing city**

| Year | Fractal dimension | | | Observed spatial entropy | | | Predicted spatial entropy | | |
|------|---------|---------|---------|---------|---------|---------|---------|---------|---------|
|      | $D_0$ | $D_1$ | $D_2$ | $M_0$ | $M_1$ | $M_2$ | $M_0$ | $M_1$ | $M_2$ |
| 1984 | 1.6932 | 1.6048 | 1.5682 | 11.7364 | 11.1236 | 10.8699 | 11.5013 | 10.8805 | 10.6373 |
| 1988 | 1.6984 | 1.6228 | 1.5967 | 11.7724 | 11.2484 | 11.0675 | 11.9127 | 11.3605 | 11.1394 |
| 1989 | 1.7491 | 1.6724 | 1.6382 | 12.1238 | 11.5922 | 11.3551 | 12.0056 | 11.4707 | 11.2553 |
| 1991 | 1.7574 | 1.6808 | 1.6518 | 12.1814 | 11.6504 | 11.4494 | 12.1804 | 11.6794 | 11.4756 |
| 1992 | 1.7437 | 1.6762 | 1.6517 | 12.0864 | 11.6185 | 11.4487 | 12.2623 | 11.7781 | 11.5801 |
| 1994 | 1.8011 | 1.7467 | 1.7199 | 12.4843 | 12.1072 | 11.9214 | 12.4158 | 11.9645 | 11.7781 |
| 1995 | 1.8171 | 1.7507 | 1.7226 | 12.5952 | 12.1349 | 11.9402 | 12.4876 | 12.0523 | 11.8716 |
| 1996 | 1.8066 | 1.7435 | 1.7172 | 12.5224 | 12.0850 | 11.9027 | 12.5562 | 12.1366 | 11.9617 |
| 1998 | 1.8101 | 1.7484 | 1.7231 | 12.5467 | 12.1190 | 11.9436 | 12.6842 | 12.2952 | 12.1315 |
| 1999 | 1.8336 | 1.7786 | 1.7568 | 12.7095 | 12.3283 | 12.1772 | 12.7439 | 12.3696 | 12.2114 |
| 2001 | 1.8531 | 1.8001 | 1.7787 | 12.8447 | 12.4773 | 12.3290 | 12.8550 | 12.5092 | 12.3617 |
| 2006 | 1.8877 | 1.8468 | 1.8277 | 13.0845 | 12.8010 | 12.6687 | 13.0897 | 12.8083 | 12.6859 |
| 2009 | 1.9033 | 1.8681 | 1.8519 | 13.1927 | 12.9487 | 12.8364 | 13.2048 | 12.9574 | 12.8486 |
| 2010 | 1.8959 | 1.8571 | 1.8384 | 13.1414 | 12.8724 | 12.7428 | 13.2394 | 13.0026 | 12.8980 |
| 2013 | 1.9357 | 1.9103 | 1.8997 | 13.4172 | 13.2412 | 13.1677 | 13.3331 | 13.1258 | 13.0333 |
| 2015 | 1.9346 | 1.9081 | 1.8968 | 13.4096 | 13.2259 | 13.1476 | 13.3879 | 13.1985 | 13.1134 |

**Note**: The fractal dimension values come from Huang (2019). The author of this paper calculated the entropy values, which are based on natural logarithm. The unit of the spatial entropy is nat.

Generally speaking, for different types of spatial entropy, the capacity values are different. However, as far as Beijing is concerned, the capacity values of Hartley entropy, Shannon entropy, and the second order Renyi entropy are the same as each other: 13.8629 nat. The number suggests that all the possible space will be occupied by built-up land. This is an unsustainable growth mode with no remaining land for future. The mode is reflected by both the changing trends of fractal dimension and spatial entropy. As indicated above, the data were extracted by using functional box-counting method. Therefore, both the fractal dimension and spatial entropy are calculated on the basis of box-counting method. The capacity value of spatial entropy depends on the spatial sphere of the largest box and its levels. The box was divided into $m=11$ levels in terms of multiple of 2. Accordingly, the number of the smallest boxes is

$$N_m = 2^{m-1} \times 2^{m-1} = 2^{2(m-1)} = 2^{20}. \tag{36}$$

Thus the maximum value of the spatial entropy is

$$M_{\max} = \ln(N_m) = \ln(2^{20}) = 20\ln(D_{\max}), \qquad (37)$$

where $D_{\max}$ denotes the maximum fractal dimension. If all the space will be made use, the maximum fractal dimension is

$$D_{\max} = d_E = 2. \qquad (38)$$

This is to say, supposing the space is completely filled, the maximum fractal dimension will equal the Euclidean dimension of the embedding space, $d_E=2$.

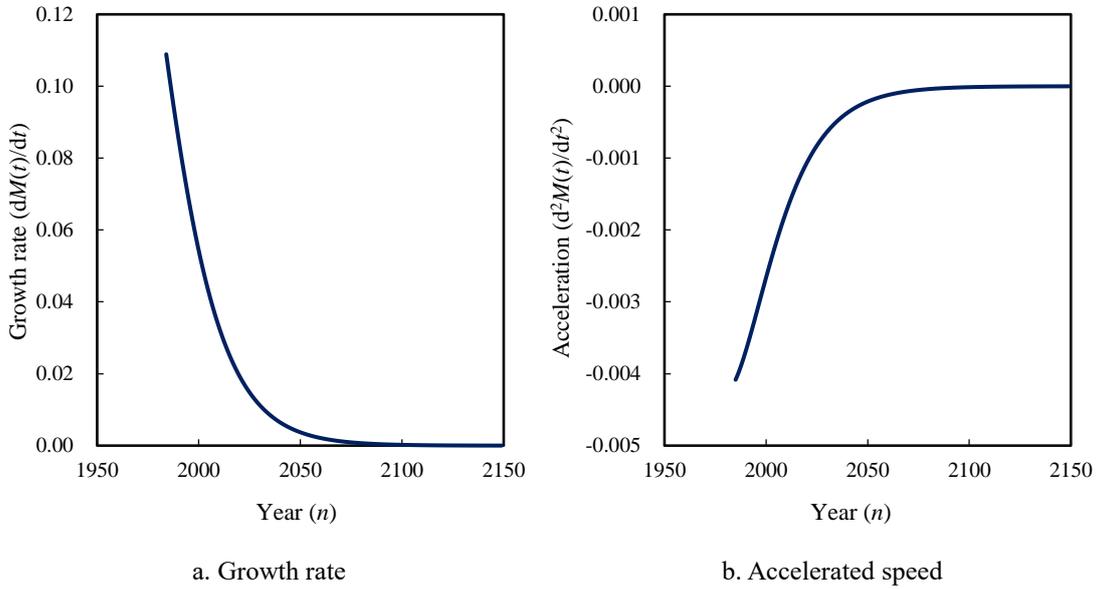

a. Growth rate  b. Accelerated speed

**Figure 3 The trend curves of growth rate and accelerated speed of spatial entropy increase of Beijing's urban form (1984-2150)**

**Note**: In order to show the change trends of growth rate and accelerated speed of spatial entropy of Beijing's urban form, the time span was extend from 1984-2015 to 1984-2050.

Using the logistic model of spatial entropy increase, we can investigate the development stages of urban growth. The first derivative of the logistic function describes the rate of increase of spatial entropy. The growth rate curve is a unimodal curve, and the peak value represents the largest value of growth rate. Corresponding to the peak, the entropy value is $M(t)=M_{\max}/2=6.9315$ nat. The second derivative of the logistic function describes the acceleration of spatial entropy. The accelerated speed curve is an S-shaped curve, which give the largest value and smallest value of the accelerated velocity. The entropy corresponding to the peak of acceleration is $M(t)= (1/2-3^{0.5}/6)M_{\max}= 2.9296$ nat, and the entropy value corresponding to the valley of acceleration is $M(t)= (1/2+3^{0.5}/6)M_{\max}=$ 10.9333 nat. By using the three special points, the logistic curve can be divided into four stages: *initial stage*, *acceleration stage*, *deceleration stage*, and *terminal stage*. The growth rate curve of Beijing's spatial entropy is a monotone decreasing curve, while the corresponding acceleration curve is a monotone increasing curve. Through trend extrapolation, the trend line based on predicted values can be extended to 2050. In this way, we can examine the possible future trends from a larger time scale. As shown, there are neither peaks nor valleys in the curves (Figure 3). In 1984, the Hartley entropy is about 11.7364 nat, which is greater than the third threshold value 10.9333 nat. This suggests that Beijing's urban growth has reached its terminal stage.

**3.2 Characterizing spatial complexity of urban form**

The spatial entropy can be transformed into corresponding spatial indexes of urban development. Suppose that the city of Beijing evolved from a point to a complex spatial system. Thus the minimum entropy is $M_{max}$ =ln(1)=0 nat. The maximum spatial entropy is $M_{max}$ =13.8629 nat, as indicated above. Using the formula displayed in Table 2, we can calculated various spatial entropy indexes. For simplicity, only Hartley entropy is taken into account. The results are as below (Table 4). This is a special case due to $M_{min}$ =ln(1)=0 nat and $M_{max}$ =13.8629 nat. The first two indexes contain no new information. Thus spatial entropy range is equal to the maximum spatial entropy, $M_{max}$ =13.8629 nat, and the space-filling index is equal to the Hartley entropy, $M(t)$. The third index, spatial information gain, is equivalent to the fifth index, spatial redundancy. The spatial redundancy multiplied by the maximum entropy yields spatial information gain. The fourth index, spatial entropy ratio, is equivalent to capacity dimension, that is, capacity dimension multiplied by 2 yields spatial entropy ratio. The normalized capacity dimension proved to be equivalent to Hartley entropy (Chen, 2020). If $M_{min}$ >0 nat and $M_{max}$ <13.8629 nat, many equivalence relationships may break. The case of Beijing can also be used to reveal the similarities and differences between the entropy indexes and fractal dimension indexes (Table 5). But this is not the subject of this article.

**Table 4 Spatial measurements based on Hartley entropy of Beijing's urban form (1984-2015)**

| Year | Spatial entropy range | Space-filling index | Spatial information gain | Spatial entropy ratio | Spatial redundancy | Spatial entropy odds |
|---|---|---|---|---|---|---|
| **1984** | 13.8629 | 11.7364 | 2.1266 | 0.8466 | 0.1534 | 5.5189 |
| **1988** | 13.8629 | 11.7724 | 2.0905 | 0.8492 | 0.1508 | 5.6313 |

| 1989 | 13.8629 | 12.1238 | 1.7391 | 0.8746 | 0.1255 | 6.9713 |
| 1991 | 13.8629 | 12.1814 | 1.6816 | 0.8787 | 0.1213 | 7.2440 |
| 1992 | 13.8629 | 12.0864 | 1.7765 | 0.8719 | 0.1282 | 6.8034 |
| 1994 | 13.8629 | 12.4843 | 1.3787 | 0.9006 | 0.0995 | 9.0553 |
| 1995 | 13.8629 | 12.5952 | 1.2678 | 0.9086 | 0.0915 | 9.9349 |
| 1996 | 13.8629 | 12.5224 | 1.3405 | 0.9033 | 0.0967 | 9.3413 |
| 1998 | 13.8629 | 12.5467 | 1.3163 | 0.9051 | 0.0950 | 9.5319 |
| 1999 | 13.8629 | 12.7095 | 1.1534 | 0.9168 | 0.0832 | 11.0192 |
| 2001 | 13.8629 | 12.8447 | 1.0182 | 0.9266 | 0.0735 | 12.6147 |
| 2006 | 13.8629 | 13.0845 | 0.7784 | 0.9439 | 0.0562 | 16.8094 |
| 2009 | 13.8629 | 13.1927 | 0.6703 | 0.9517 | 0.0484 | 19.6825 |
| 2010 | 13.8629 | 13.1414 | 0.7216 | 0.9480 | 0.0521 | 18.2123 |
| 2013 | 13.8629 | 13.4172 | 0.4457 | 0.9679 | 0.0321 | 30.1042 |
| 2015 | 13.8629 | 13.4096 | 0.4533 | 0.9673 | 0.0327 | 29.5810 |

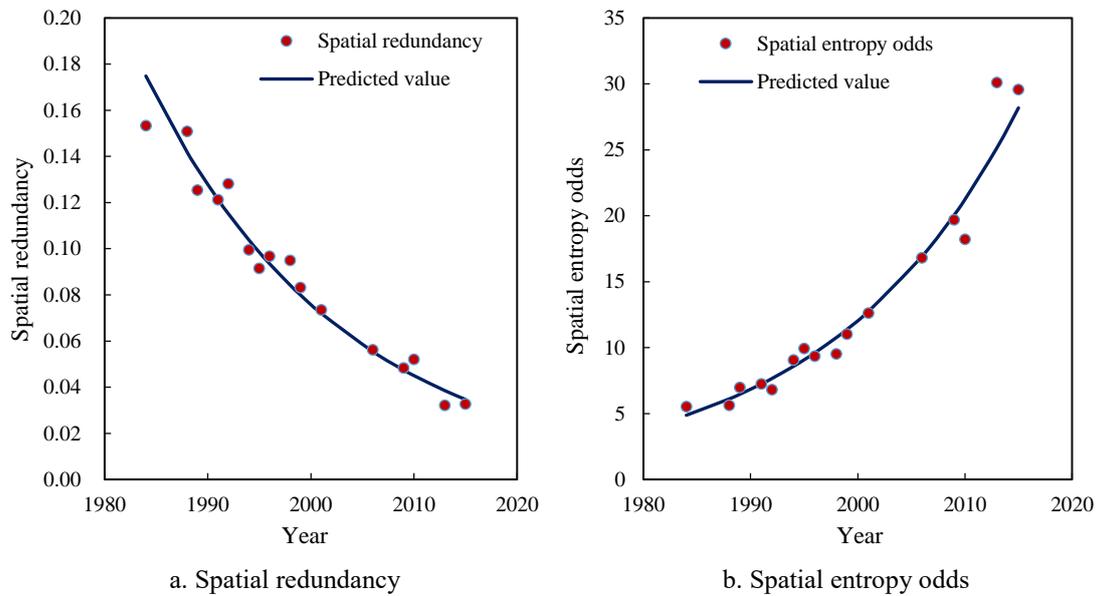

a. Spatial redundancy        b. Spatial entropy odds

**Figure 4 The spatial redundancy and spatial entropy odds of Beijing's urban form and growth**

**(1984-2015)**

Because the urban growth of Beijing has reached the terminal stage and its growing pattern is special, it is relatively easy to describe the spatial complexity of Beijing's urban development. In this case, two spatial measurements are enough to characterize the spatial complexity of Beijing's urban form and growth. The two indexes are spatial redundancy and spatial entropy odds (Figure 4). First, let's examine the spatial redundancy, which can also be termed space-saving ratio. The spatial redundancy takes on an exponential decay over time. The exponent decay model is

$$R_y(t) = 0.1749e^{-0.0522t}, \quad (39)$$

where $t=n-1984$, and $n$ represent year. The goodness of fit is $R^2=0.9682$, and the characteristic length of time is about $1/0.0522 \approx 19$ year. The time corresponds to 2002, and 2002 minus 1984 equals 19. This suggests that, after 2002, the spatial redundancy will rapidly become no significant difference from 0. If the spatial redundancy approaches to zero, then there is not much land available for Beijing's development. Next, let's examine the spatial entropy odds. The spatial entropy odds takes on an exponential increase over time. The exponential growth model is

$$O(t) = 4.8700e^{0.0566t}. \quad (40)$$

The goodness of fit is $R^2=0.9709$. It can be demonstrated that there is mathematical transformation relation between equation (40) and equation (33). The characteristic length of time is about $1/0.0566 \approx 18$ year. The time corresponds to 2001, and 2001 minus 1984 equals 18. This suggests that after 2001, the spatial entropy odds will approach the limit rapidly. If $M_{max} < 13.8629$ nat, the spatial entropy odds will follows logistic function rather than exponential function. The exponential growth model lends further support to the judgment that the land use of Beijing bears an unsustainable growth mode with no remaining land for future.

**Table 5 Spatial measurement results based on spatial entropy and fractal dimension of Beijing's urban form (2015)**

| Spatial indexes based on fractal dimension | | | Spatial indexes based on entropy | | |
|---|---|---|---|---|---|
| Measurement | Formula | Result | Measurement | Formula | Result |
| Fractal dimension range | $D_{max} - D_{min}$ | 2 | Spatial entropy range | $M_{max} - M_{min}$ | 13.8629 |
| Space-filling degree | $D(t) - D_{min}$ | 1.9346 | Space-filling index | $M(t) - M_{min}$ | 13.4096 |
| Space-saving degree | $D_{max} - D(t)$ | 0.0654 | Spatial information gain | $M_{max} - M(t)$ | 0.4533 |
| Space-filling ratio | $\dfrac{D(t) - D_{min}}{D_{max} - D_{min}}$ | 0.9673 | Spatial entropy ratio | $\dfrac{M(t) - M_{min}}{M_{max} - M_{min}}$ | 0.9673 |
| Space-saving ratio | $\dfrac{D_{max} - D(t)}{D_{max} - D_{min}}$ | 0.0327 | Spatial Redundancy | $\dfrac{M_{max} - M(t)}{M_{max} - M_{min}}$ | 0.0327 |

| Fractal dimension odd | $\dfrac{D(t)}{D_{\max} - D(t)}$ | 29.5810 | Spatial entropy odds | $\dfrac{M(t)}{M_{\max} - M(t)}$ | 29.5810 |

**Note**: For the first three calculation results, the index value based on entropy is 6.9315 times of that based on fractal dimension (ln($2^{10}$)=6.9315); for the last three calculation results, the index value based on entropy is equal to the index value based on fractal dimension.

## 4. Discussion

According to the theoretical derivation, we can use logistic function and Boltzmann equation to model spatial entropy increase of urban development. Based on the logistic model and Boltzmann model, we can construct a set of entropy indexes to characterize urban form. To calculate spatial entropy, we have to make use of zonal systems (Figure 1). In practice, a zonal system of cities always form a hierarchy with cascade structure (Batty, 1974; Batty, 1976). The hierarchy based on twofold cascading can be treated as spatial disaggregation (Batty and Longley, 1994). In light of ideas from fractals, we can understand the cascade structure from a new angle of view. In fact, a city can be divided into several sectors, a sector can be divided into several districts, a district can be divided into several neighborhoods, and a neighborhood can be divided into several sites (Kaye, 1989). In each part of each level, we can find residential land, commercial-industrial land, vacant land, and open space, and so on. Different levels of different parts form a hierarchy with cascade structure, which indicates a zonal system with self-similar process. We can use fractal dimension to describe the self-similar patterns, or use spatial entropy to characterize the spatial structure. The spheres of application of the models and measurements presented above lie in two aspects: artificially defined zonal system, e.g., urban district system, the systems with characteristic lengths, e.g., the zonal systems for population density distributions.

In fact, multifractal dimension spectrums are based on Renyi entropy, and in turn we can derive generalized entropy spectrum for urban studies. A preliminary model has been proposed in previous work (Chen and Feng, 2017). Based on box-counting method, normalized spatial entropy proved to equal normalized fractal dimension (Chen, 2020). If $M_{\min}$=0, then we have $D_{\min}$=0. For a non-fractal system or the non-fractal aspect of a fractal system, $D_{\max}$=d=2. Then, based on equation (28), a pair of generalized global multifractal parameters can be defined as below

$$D_q^* = \frac{2}{M_{max}} M_q, \tag{41}$$

$$\tau_q^* = \frac{2}{M_{max}}(q-1)M_q, \tag{42}$$

where $D_q^*$ denotes the generalized correlation dimension in the broad sense, and $\tau_q^*$ refers to the generalized mass exponent. Then, by using the Legendre transform, we can derive a pair of generalized local multifractal parameters. One is the generalized singularity exponent $\alpha^*(q)$, and the other, is the corresponding generalized local fractal dimension $f(\alpha^*)$. The two indicators can be calculated by

$$\alpha^*(q) = \frac{d\tau_q^*}{dq} = \frac{2}{M_{max}}(M_q + (q-1)\frac{dM_q}{dq}), \tag{43}$$

$$\begin{aligned} f(\alpha^*(q)) &= q\alpha^*(q) - \tau_q^* = q\alpha^*(q) - \frac{2}{M_{max}}(q-1)M_q \\ &= \frac{2}{M_{max}}(M_q + q(q-1)\frac{dM_q}{dq}) \end{aligned}. \tag{44}$$

Using equations (41), (42), (43), and (44), we can transform the global generalized entropy measurements into the local generalized entropy measurements and *vise versa*.

Multifractal theory is a good mathematical tools for urban studies. However, the standard multifractal parameters can only be utilized to analyze complex spatial systems with multi-scaling processes and patterns. In contrast, the generalized multifractal indexes can be applied to both fractal systems and non-fractal systems, and the functional box-counting methods can be replaced with zonal systems in practice. Anyway, a fractal dimension can only be measured for fractal phenomena, but spatial entropy can be measured for any type of geographical systems. The general formulae, equations (41) and (43), are on the basis of spatial entropy rather than fractal dimension. By analogy, a new framework of spatial analysis can be developed on the basis of the generalized multifractal parameters (Table 6).

**Table 6 The correspondence relationships between spatial entropy spectrums and multifractal dimension spectrums for spatial analysis of cities**

| Parameter level | | Spatial entropy | Fractal dimension |
|---|---|---|---|
| **Global** | General | Renyi entropy spectrum, $M_q$ v.s. $q$ | Global multifractal dimension |
| parameter | | | spectrum, $D_q$ v.s. $q$ |
| | Special | Hartley entropy, $M_0$ | Capacity dimension, $D_0$ |
| | | Shannon entropy, $M_1$ | Information dimension, $D_1$ |
| | | The 2nd order Renyi entropy, $M_2$ | Correlation dimension, $D_2$ |
| Connection | | Legendre transform | |
| Local parameter | General | Generalized Shannon entropy spectrum, $f(\alpha^*)$ v.s. $\alpha^*$ | Local multifractal dimension spectrum, $f(\alpha)$ v.s. $\alpha$ |
| | | Mixed entropy, $\alpha^*(q)$ | Singularity exponent, $\alpha(q)$ |
| | Special | Hartley entropy, $M_0$ | Capacity dimension, $D_0$ |
| | | Shannon entropy, $M_1$ | Information dimension, $D_1$ |

In fractal dimension modeling of urban growth, the logistic function and Boltzmann equation can be generalized to quadratic logistic function and quadratic Boltzmann equation. Accordingly, in spatial entropy modeling of urban development, we have corresponding quadratic models. The quadratic logistic function of spatial entropy increase is

$$M_q(t) = \frac{M_{max}}{1+(M_{max}/M_q(0)-1)e^{-rt^2}}, \qquad (45)$$

which can be easily derived from the quadratic logistic model of fractal dimension increase of urban form. The quadratic Boltzmann equation of spatial entropy increase is

$$M_q(t) = M_{min} + \frac{M_{max}-M_{min}}{1+(\frac{M_{max}-M_q(0)}{M_q(0)-M_{min}})e^{-(rt)^2}} = M_{min} + \frac{M_{max}-M_{min}}{1+\exp(-\frac{t^2-\tau^2}{p^2})}, \qquad (46)$$

where the parameters $\tau^2=p^2\ln((D_{max}-D_q(0))/(D_q(0)-D_{min}))$, $p^2=1/r^2$. If the minimum entropy $M_{min}=0$, the quadratic Boltzmann equation will become quadratic logistic function. Equation (46) can be readily derived from the quadratic Boltzmann model of fractal dimension increase of urban form.

The urban development and evolution of Beijing has aroused the research interest of many scholars. A great many research results on Beijing's urban growth emerged in the last 20 years. The main studies includes four aspects as below: CA-(cellular automata-) based simulation (e.g., Chen *et al*, 2002; Deng *et al*, 2019; Guan *et al*, 2005; Long *et al*, 2013), urban sprawl (e.g., Jiang *et al*, 2007; Ma *et al*, 2008; Zhao *et al*, 2010), urban land use (e.g., Deng and Srinivasan, 2016; Liu *et al*, 2002), and city planning (e.g.. Wang *et al*, 2020; Yu *et al*, 2011; Zhao and Lü, 2011; Zhao *et al*, 2011). Among various interesting studies, few works are devoted to mathematical modeling of the growing process of Beijing. In recent years, the ideas from growing fractals and multifractals were employed to explore the urban form and growth of Beijing (e.g., Chen and Huang, 2019; Chen and Wang, 2013). Fractal dimension is defined on the basis of entropy, and fractal dimension models of cities can be generalized to spatial entropy modeling in principle. This work is devoted to developing spatial measurements and models of urban growth and form. Compared with the previous studies, the potential contributions of this study are as follows. First, the models of spatial entropy increase of urban growth are derived from the fractal dimension curve models. Second, a set of spatial measurements are constructed by analogy with the related fractal parameters. Third, the relationships between spatial entropy and fractal dimension are revealed and the generalized multifractal indicators are improved. Mathematical modeling is important for scientific studies of cities. As Neumann (1961) pointed out: "The sciences do not try to explain, they hardly even try to interpret, they mainly make models." In fact, scientific method includes two elements: description and understanding. Scientific research always proceed first by describing how a system and its parts work and later by understanding why (Gordon, 2005; Henry, 2002). Precise descriptions and determinations of a system rely heavily on mathematical modeling and measurements, while understanding the work mechanism of a system depends on observations, experience, laboratory experiments, and computer simulation (Bak, 1996; Einstein, 1953; Henry, 2002; Waldrop, 1992). The models and measurements proposed in this paper are useful for descriptions and predictions of urban growth and understanding spatial complexity of cities.

The models of fractal dimension increase curves are only suitable for the urban form without characteristic scale. Compared with the fractal-based models of urban growth, the entropy-based models of urban growth have significant advantages: First, it can be used to precisely describe both the urban growth with characteristic scales and that without characteristic scales. Second, it has a potential function for understanding spatial complexity, in particular, spatial dynamics, of urban evolution. The disadvantage lies in that the determinations of spatial entropy depend on measurement scales. Moreover, the shortcomings of this study lies in two aspects. On the one hand, the quadratic logistic function and Boltzmann equation are less discussed. On the other hand, the influence of the modifiable areal unit problem (MAUP) on spatial entropy are not taken into consideration. The problems remain to be solved in future studies.

# 5. Conclusions

Entropy and fractal dimension have both intersectional domain and different spheres of application to urban research. The key lies in the equivalence relation between normalized entropy and normalized fractal dimension. For a given spatial measurement scale, fractal dimension can often be replaced by entropy for simplicity. The models and measurements of this paper are chiefly suitable for artificially defined zonal systems and the systems with characteristic scales. The main points of this work can be outlined as follows. **First, the spatial entropy increase of urban growth can be modeled by logistic function and Boltzmann equation.** This is the first common point of spatial entropy and fractal dimension. The fractal dimension growth curves can be modeled by both logistic function and Boltzmann equation. By means of the equivalence relation between normalized entropy and normalized fractal dimension, we can derive the models of spatial entropy increase from the models of fractal dimension increase of urban form and *vice versa*. **Second, based on normalized entropy, a set of spatial measurements can be constructed to describe the space filling, spatial uniformity, and spatial complexity.** This is the second common point of spatial entropy and fractal dimension. Based on normalized fractal dimension, a series of fractal indexes can be constructed to characterize urban growth and form. We can derive spatial entropy indexes from fractal indexes and *vice versa* by means of the transformation relations between normalized entropy and normalized fractal dimension. **Third, the difference between the models and measurements of spatial entropy increase and those of fractal dimension rests with characteristic length.** Fractal dimension is only suitable for describing the complex spatial systems with scaling nature. If and only if a system bear no characteristic scales, we can adopt fractal geometry to characterize it. If a system possesses characteristic scales, it will has no fractal dimension. The application scope of spatial entropy is wider than that of fractal dimension. If some aspects of a city or urban systems have characteristic scales, the fractal dimension cannot be measured. In this case, we can only use spatial entropy rather than fractal dimension to describe the spatial features of urban evolution.

# Acknowledgements

This research was sponsored by the National Natural Science Foundation of China (Grant No.

41671167). Dr. Linshan Huang of Peking University for providing spatial datasets of Beijing's land use. The two anonymous reviewers whose constructive comments were helpful in improving the quality of this paper. The support and help are gratefully acknowledged. The paper has been submitted to arXiv as a pre-print, see the following link: "https://arxiv.org/abs/2005.02272".

## Conflict of interest statement

The author declares that there is no conflict of interests regarding the publication of this article.